\documentclass[amsmath,amssymb,aps,prl,reprint,superscriptaddress,longbibliography]{revtex4-2}

\usepackage{graphicx}
\usepackage{xcolor}
\definecolor{mylinkcolor}{rgb}{0,0,0.64}
\usepackage{hyperref}
\hypersetup{colorlinks=true, breaklinks=true, urlcolor=mylinkcolor, linkcolor=mylinkcolor, citecolor=mylinkcolor}

\begin{document}


\title{Spin-Orbit Coupling-Driven Chirality Switching of Spin Waves in Altermagnets}


\newcommand{\IOP}{Beijing National Laboratory for Condensed Matter Physics and Institute of Physics, Chinese Academy of Sciences, Beijing 100190, China}
\newcommand{\UCAS}{School of Physical Sciences, University of Chinese Academy of Sciences, Beijing 100049, China}
\newcommand{\SLAB}{Songshan Lake Materials Laboratory, Dongguan, Guangdong 523808, China}
\newcommand{\SYSU}{School of Physics, Sun Yat-Sen University, Guangzhou 510275, China}

\author{Wen-Tong~Li}
\affiliation{\IOP}
\affiliation{\UCAS}

\author{Yu-Biao~Wu}
\affiliation{\IOP}

\author{Lin~Zhuang}
\email{stszhl@mail.sysu.edu.cn}
\affiliation{\SYSU}

\author{Jian-Tao~Wang}
\affiliation{\IOP}
\affiliation{\UCAS}
\affiliation{\SLAB}

\author{Wu-Ming~Liu}
\email{wmliu@iphy.ac.cn}
\affiliation{\IOP}

\date{\today}

\begin{abstract}
Altermagnets host intrinsically chirality-splitting spin waves, which offer an ideal platform for chirality-based computing with low energy consumption and fast dynamics. However, achieving precise and efficient control over spin-wave chirality remains a challenge.
Here, we propose a mechanism to switch the chirality of spin waves in altermagnets via electrically induced Rashba spin-orbit coupling (SOC), which is free of tuning external fields.
For in-plane spin polarization, SOC introduces a splitting effect opposite to the altermagnetism, leading to spin inversion in the electronic energy bands and chirality reversal in the spin-wave dispersion. By tuning SOC strength, the chirality splitting of spin waves can be controllably modified, enabling chirality switching at fixed resonance conditions, which results in the reversal of transverse spin susceptibility.
We further design an experimental setup based on an altermagnet/antiferromagnet heterostructure to realize this mechanism.
Our work establish a pathway toward efficient electrical control of spin-wave chirality in altermagnets, facilitating the development of chirality-based spintronic devices.
\end{abstract}


\maketitle


\textit{Introduction---}\noindent
Chiral spin waves (or their quanta magnons) in magnetic materials present prospective information carriers for chirality-based computing \cite{chumak_Magnon_2015, neusser_Magnonics_2009, mahmoud_Introduction_2020}, offering advantages such as low energy consumption and fast dynamics \cite{che_Discovery_2025, takeuchi_Electrical_2025, leenders_Canted_2024a} compared to charge and spin.
Conventional magnets, however, face limitations for chirality operations: ferromagnets support only right-handed (RH) modes, while in antiferromagnets the RH and left-handed (LH) chiralities remain fully degenerate \cite{smejkal_Chiral_2023}.
The recent emergence of altermagnets, which are identified as a third class of collinear magnets within the nonrelativistic spin group theory \cite{smejkal_Conventional_2022, smejkal_Emerging_2022}, provides a promising alternative.
These materials \cite{krempasky_Altermagnetic_2024, lee_Broken_2024, reimers_Direct_2024, zeng_Observation_2024, jiang_Metallic_2025} combine spin-splitting features akin to ferromagnets with zero net magnetization characteristic of antiferromagnets \cite{smejkal_Conventional_2022, smejkal_Emerging_2022, krempasky_Altermagnetic_2024, lee_Broken_2024, reimers_Direct_2024, zeng_Observation_2024, jiang_Metallic_2025, das_Realizing_2024, liao_Separation_2024a, gu_Ferroelectric_2025}, leading to novel phenomena \cite{song_Altermagnets_2025} in charge-spin conversion \cite{gonzalez-hernandez_Efficient_2021a, zhang_Simultaneous_2024, guo_Direct_2024}, magneto-optical effects \cite{zhou_Crystal_2021, han_Electrical_2024}, tunneling magnetoresistance \cite{smejkal_Giant_2022, liu_Giant_2024, noh_Tunneling_2025} and anomalous thermal transport \cite{zhou_Crystal_2024}.
Crucially, altermagnets host intrinsic chirality-splitting spin waves \cite{das_Anisotropic_2022, smejkal_Chiral_2023, cui_Efficient_2023, hodt_Spin_2024, liu_Chiral_2024}, making them a highly suitable platform for implementing chirality-based information processing.

A central challenge in achieving practical chiral computing lies in the precise control of chirality.
Current experimental methods for chirality switching predominantly rely on extrinsic parameters \cite{wang_Perspectives_2024, sheng_Control_2025, liu_Switching_2022} such as temperature, excitation frequency or external magnetic fields to selectively activate specific chiral modes.
However, these approaches suffer from high energy costs, slow switching speeds, or limited compatibility with integrated device architectures.
Electrical control \cite{takeuchi_Electrical_2025, zhang_Predictable_2024, han_Electrical_2024} represents a compelling alternative to address these shortcomings. While recent studies have demonstrated electrical manipulation of the spin-splitting structure in altermagnets \cite{duan_Antiferroelectric_2025, chen_Electrical_2025, gu_Ferroelectric_2025}, achieving direct electrical switching of spin-wave chirality has remained unexplored. The Rashba spin-orbit coupling (SOC), as an intrinsic and electrically tunable mechanism \cite{manchon_Currentinduced_2019, wadley_Electrical_2016, guo_Firstorder_2021}, plays a critical role in influencing magnetic anisotropy \cite{calder_Spinorbitdriven_2016, gui_Geometric_2019} and governing spin relaxation and transport dynamics \cite{sinova_Spin_2015, sun_Classification_2016, manchon_Currentinduced_2019}. It thus offers a promising pathway for enabling efficient and scalable chirality control.

In this Letter, we propose a mechanism for reversible switching of chiral spin-wave modes in altermagnets by tuning SOC strength, without changing external fields.
Employing nonequilibrium quantum field theory, we derive the effective action for a d-wave altermagnet with induced Rashba SOC to analyze the magnetic properties.
For in-plane spin polarization, spin inversion of the electronic bands and chirality reversal of the spin-wave dispersion both reveal that SOC introduces opposite spin-splitting and chirality-splitting effects against the altermagnetism.
Leveraging the competition between SOC and altermagnetism, the splitting structure of chiral spin-wave modes can be reversed by tuning the intrinsic parameter of the system.
We also design an experiment based on the altermagnet/antiferromagnet heterostructure to realize the proposed chirality-switching mechanism.
Our results establish SOC as a powerful tool for electrically manipulating chiral excitations in altermagnets and provide controlled switching of chirality bits for chirality-based spintronic devices.

\begin{figure}[htbp]
	\includegraphics[width=\columnwidth]{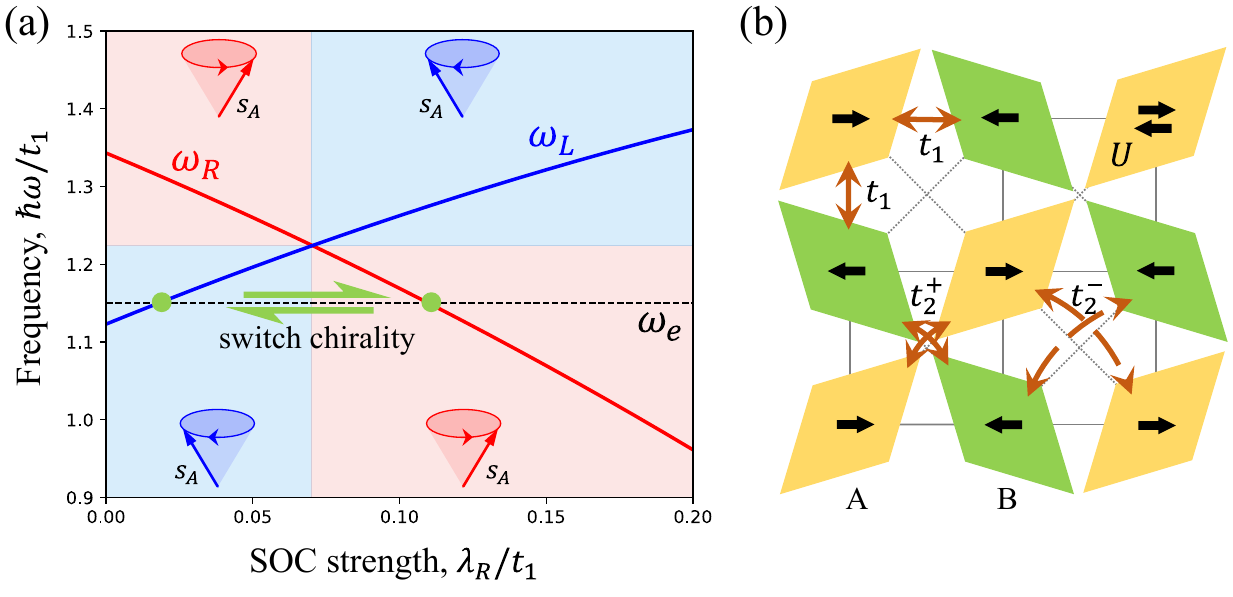}
	\caption{(a) Schematic illustration of SOC-driven chirality switching. The red and blue curves represent the RH ($\omega_R$) and LH ($\omega_L$) spin-wave modes, respectively. At the fixed resonance frequency ($\omega_e$), the chirality can be switched between different modes by tuning SOC strength. The RH and LH modes are defined by the precessing direction of sublattice spins $s_{A}=-s_B$. (b) Altermagnetic lattice configuration. The nearest-neighbor hopping $t_1$, next-nearest-neighbor hopping $t_2^\pm=t'(1\pm\delta)$ and on-site interaction $U$ are included.}
	\label{pic_model}
\end{figure}

\textit{Effective action of the altermagnet with SOC---}\noindent
The RH and LH spin waves are split in altermagnets due to the special crystallographic symmetry.
Our principle for chirality switching is to introduce an additional splitting mechanism that competes with the altermagnetism, thereby modulating the spin-wave dispersion and enabling chirality switching without varying the excitation field.
Remarkably, we find that Rashba SOC serves precisely this purpose. 
As illustrated in Fig.~\ref{pic_model}(a), increasing the SOC strength reverses the chirality-splitting structure. At a fixed resonance frequency, the dominant mode switches from LH to RH---a transition that is reversible upon reducing the SOC strength.

To elucidate this mechanism, we consider an altermagnetic film coupled to an antiferromagnetic substrate, where broken structural inversion symmetry in the latter induces a Rashba SOC \cite{zelezny_Relativistic_2014, manchon_Currentinduced_2019}. The system is described by the Hamiltonian $\mathcal{H}=\mathcal{H}_0+\mathcal{H}_I$, with
\begin{align}
	&\mathcal{H}_0=-t_1\sum_{\langle ij\rangle}(a_i^\dagger b_j+\mathrm{h.c.})-\sum_{\langle\langle ij\rangle\rangle}(t_2^A a_i^\dagger a_j+t_2^B b_i^\dagger b_j) \nonumber\\
	&\qquad\,\,\,+i\lambda_R\sum_{\langle ij\rangle}\big[a_i^\dagger (\sigma^x d_{ij}^y-\sigma^y d_{ij}^x) b_j+\mathrm{h.c.}\big], \nonumber\\
	&\mathcal{H}_I=U\sum_i(n^A_{i\uparrow}n^A_{i\downarrow}+n^B_{i\uparrow}n^B_{i\downarrow}), \label{H}
\end{align}
where $a$ and $b$ annihilate electrons on sublattice A and B, respectively.
$t_1$ is the nearest-neighbor hopping integral and taken as the energy unit. The diagonal hopping integrals are anisotropic: in the $\pm(1,1)$ directions $t_2^A=t'(1+\delta)$, $t_2^B=t'(1-\delta)$ and in the $\pm(1,-1)$ directions they are interchanged, as shown in Fig.~\ref{pic_model}(b). Such a configuration makes different sublattices connected by $C_{4z}\mathcal{T}$ symmetry, which is the characteristic of altermagnets. $\lambda_R$ is the SOC strength. $\sigma^\mu$ are Pauli matrices in the spin space, and $\mathbf{d}_{ij}$ represents the nearest-neighbor connecting vector from lattice $j$ to $i$. $U$ is the repulsive Hubbard interaction.

The non-interacting part in terms of band basis $\Psi_\mathbf{k}=(a_{\mathbf{k}\uparrow},b_{\mathbf{k}\uparrow},a_{\mathbf{k}\downarrow},b_{\mathbf{k}\downarrow})^T$ in the momentum space reads $\mathcal{H}_0=\sum_\mathbf{k}\Psi_\mathbf{k}^\dagger h_0(\mathbf{k})\Psi_\mathbf{k}$, where $h_0(\mathbf{k})=\varepsilon_0(\mathbf{k})\Gamma_{00}+\varepsilon_1(\mathbf{k})\Gamma_{01}+\varepsilon_2(\mathbf{k})\Gamma_{03}+\varepsilon_3(\mathbf{k})\Gamma_{11}+\varepsilon_4(\mathbf{k})\Gamma_{21}$.
$\Gamma_{\mu\nu}=\sigma^\mu\otimes\tau^\nu$ are the combined Pauli matrices acting on the spin and sublattice spaces.
The band parameters are $\varepsilon_0(\mathbf{k})=-4t'\cos k_x\cos k_y$, $\varepsilon_1(\mathbf{k})=-2t_1(\cos k_x+\cos k_y)$, $\varepsilon_2(\mathbf{k})=4t'\delta\sin k_x\sin k_y$, $\varepsilon_3(\mathbf{k})=2\lambda_R\sin k_y$, $\varepsilon_4(\mathbf{k})=-2\lambda_R\sin k_x$.
The anisotropy $\delta$ manifests itself in $\varepsilon_2$ with $C_{4z}\mathcal{T}$ symmetry, while SOC terms $\varepsilon_{3,4}$ have $C_{2y}$ and $C_{2x}$, respectively.
Their distinct symmetry underlies competing effects.

Through Hubbard-Stratonovich transformation, the interaction is decoupled by introducing bosonic magnetization fields $\mathbf{m}=\mathbf{m}_A=-\mathbf{m}_B$ on each sublattice \cite{ishiwata_Axion_2021}. 
Expressing all terms in the Keldysh formalism and integrating out the electrons (details in S1 of Supplemental Material \cite{SM}), we obtain an effective action
\begin{equation}
	\mathcal{S}_{\rm eff}=-i\operatorname{Tr}\ln[\check{G}^{-1}+\check{M}]-\dfrac{12}U\int dx\,\mathbf{m}_c(x)\cdot\mathbf{m}_q(x), \label{action}
\end{equation}
where $\check{G}$ represents the dressed Green's function matrix of electrons, and $\check{M}=(\mathbf{m}_c\gamma^0+\mathbf{m}_q\gamma^1)\otimes\Gamma_{\mu\nu}$ has the meaning of the self-energy matrix, with $\gamma^\mu$ Pauli matrices in the Keldysh space.
$c\,(q)$ denotes the classical (quantum) Keldysh component.
This action describes the static and dynamic magnetic properties of the system.

\textit{Spin Inversion in Electronic Bands---}\noindent
The equilibrium state is supposed to satisfy the stationary condition $\frac{\delta \mathcal{S}_{eff}}{\delta m_q}\big|_{m_q=0}=0$, which yields the self-consistent equation of the magnetic order parameter
\begin{equation}
	\langle m_0^i\rangle=-i\frac U{12}\operatorname{Tr}[\check{G}\gamma^1\otimes\Gamma_{i3}]. \label{order parameter}
\end{equation}
For simplicity, we assume $\mathbf{m}_0$ is a constant and perform a numerical calculation at half filling (details in S2 of Supplemental Material \cite{SM}).
The mean-field interacting Hamiltonian can be extracted from the dressed Green's function as $h(\mathbf{k})=h_0(\mathbf{k})-m_0^i\Gamma_{i3}$. Diagonalizing it gives the electronic bands
\begin{align}
	&E(\mathbf{k})=\varepsilon_0\pm\sqrt{|\varepsilon|^2+m_0^2\pm2\sqrt{\varepsilon_1^2(\varepsilon_3^2+\varepsilon_4^2)+\Delta_{cp}}}, \nonumber \\
	&\Delta_{cp}=m_0^2\varepsilon_2^2+ m_z^2(\varepsilon_3^2+\varepsilon_4^2)+m_x^2\varepsilon_4^2+m_y^2\varepsilon_3^2 \nonumber \\
	&\qquad\quad-2m_x\varepsilon_1\varepsilon_2\varepsilon_3-2m_y\varepsilon_1\varepsilon_2\varepsilon_4-2m_xm_y\varepsilon_3\varepsilon_4,
\end{align}
where $|\varepsilon|^2=\sum_{\alpha=1}^4\varepsilon_\alpha$ and $m_0^2=|\mathbf{m}_0|^2=m_x^2+m_y^2+m_z^2$.
The outer square root opens a gap between the valence and conduction bands, while the inner one lifts the intraband Kramers degeneracy.
$\Delta_{cp}$ indicates the coupling between SOC ($\varepsilon_{3,4}$) and altermagnetism ($m_0,\,\varepsilon_2$), and it has explicit dependence on the spin polarization orientation.

\begin{figure}[htbp]
	\includegraphics[width=\columnwidth]{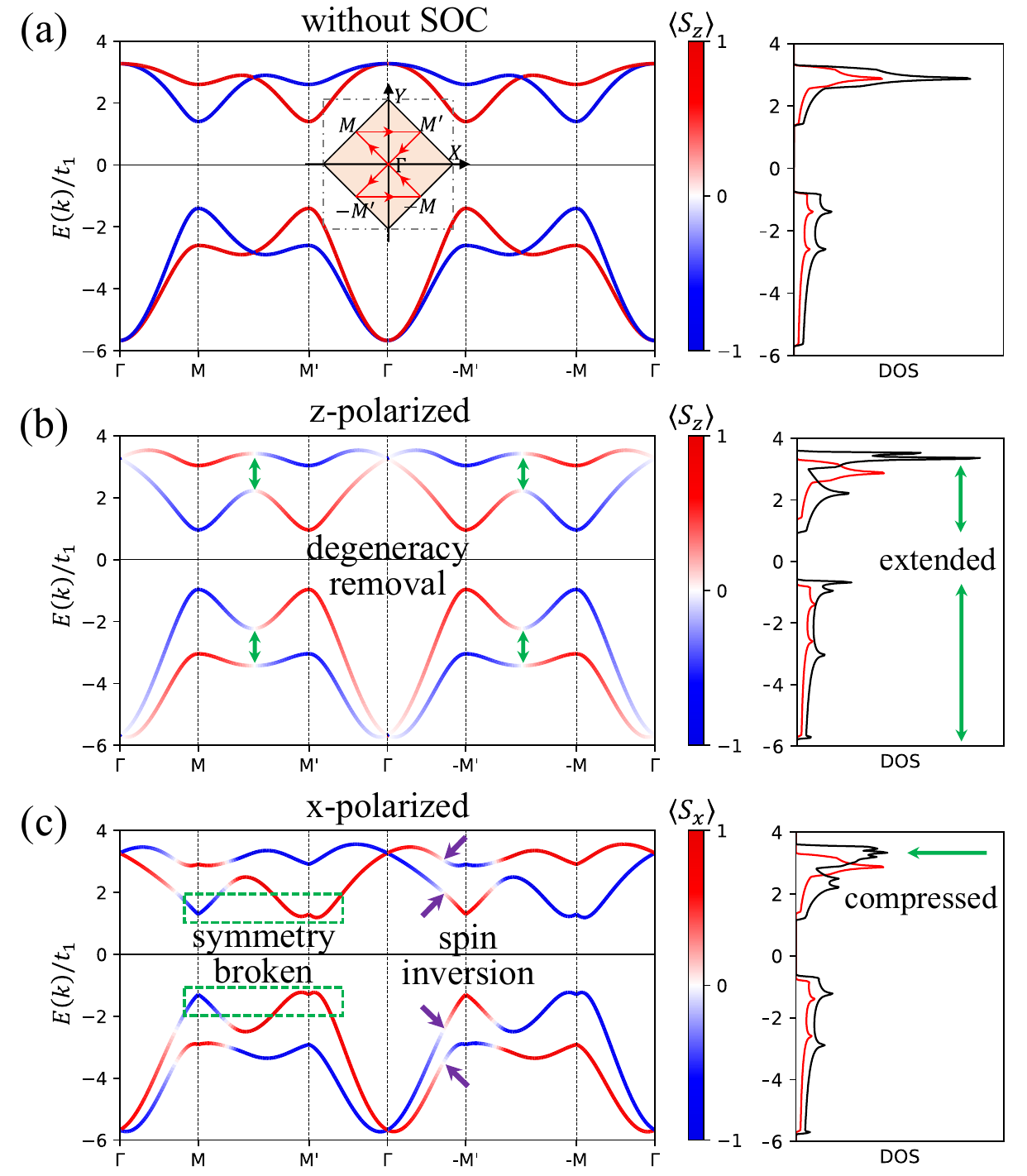}
	\caption{Spin-projected energy bands and density of states (DOS). The red lines of DOS are the spin-up component without SOC for comparison. (a) Without SOC, the bands are well polarized, and DOS is just twice of the spin-up component. The inset shows the reduced Brillouin zone and high-symmetry path. (b) With SOC and z polarization, the degeneracy on the high-symmetry paths is removed, and the $C_{4z}\mathcal{T}$ symmetry is preserved. The DOS is extended slightly. (c) With SOC and x polarization, the $C_{4z}\mathcal{T}$ symmetry is broken, and intraband spin inversion arises. The maximum value of DOS is compressed to a half due to the asymmetry between spin-up and spin-down components. The parameters are $t'=0.3,\delta=0.5,\lambda_R=0.3,m_0=2$.}
	\label{pic_band}
\end{figure}

Fig.~\ref{pic_band}(a) shows the spin-projected band structure without SOC along the high-symmetry path in the inset and corresponding density of states (DOS). The two spin components are well split due to altermagnetism.
Notably, the band structure for x polarization is identical to the z-polarizing case when SOC is off.
The total DOS is just twice of the spin-up component.
With SOC on, $\Delta_{cp}$ becomes orientation-dependent.
For z polarization, $\Delta_{cp}=m_z^2(\varepsilon_2^2+\varepsilon_3^2+\varepsilon_4^2)$. The quadratic form means SOC and altermagnetism both contribute to the intraband spin splitting, and the $C_{4z}\mathcal{T}$ symmetry is preserved.
As shown in Fig.~\ref{pic_band}(b), the band symmetry and spin projection are the same with Fig.~\ref{pic_band}(a), but a larger splitting difference between the spin components arises. The degeneracy on the high-symmetry paths $X-\Gamma-(-X)$ and $Y-\Gamma-(-Y)$, where $\varepsilon_2=0$, is also lifted.
The peak value of total DOS remains twice of the non-SOC spin-up component while has a misplacement, indicating the synchronous shift of both spin components.

However, two linear terms $-2m_x\varepsilon_1\varepsilon_2\varepsilon_3$, $-2m_y\varepsilon_1\varepsilon_2\varepsilon_4$ and a mixed term $-2m_xm_y\varepsilon_3\varepsilon_4$ arise for in-plane polarization, which contain the linear factor $\sin k_x$ or $\sin k_y$ and may be negative, implying a competing effect between SOC and altermagnetism. For simplicity, we focus on x polarization to avoid the mixing.
Since $-2m_x\varepsilon_1\varepsilon_2\varepsilon_3$ contains $\sin k_x$, the $C_{2y}$ and $C_{4z}\mathcal{T}$ symmetries are broken. When $-\pi<k_x<0$, the negative value reduces the spin splitting, as shown in $\Gamma-M$ and $\Gamma-(-M')$ ranges of Fig.~\ref{pic_band}(c). More importantly, spin inversion emerges in these ranges, indicating the crossover between two distinct spin-splitting origins.
Compared with Fig.~\ref{pic_band}(a), we find the short-wavelength range is dominated by altermagnetism, and therefore, the long-wavelength range is dominated by SOC.
The peak value of total DOS is compressed to a half, implying the asynchronous shift of two spin components.
Additionally, we also give the magnetic phase diagrams in S4 of Supplemental Material \cite{SM}, which confirm the destabilizing effect of SOC on the magnetic order.

\textit{Chirality Switching of Spin Waves---}\noindent
To describe the spin dynamics, we introduce the spin susceptibility
\begin{equation}
	\chi^{RPA}_{ij}=\chi_{ij}^0(\delta_{ij}-\dfrac U6\chi_{ij}^0)^{-1}, \label{RPA polarization}
\end{equation}
within random phase approximation (RPA), where $\chi^0_{ij}$ is the bare susceptibility and the detailed derivation is in S3 of Supplemental Material \cite{SM}.
The divergence of the spin susceptibility in an ordered phase, $\chi^{RPA,-1}_{ij}(\textbf{q},\omega)=0$, indicates resonance of the response to an external perturbation, which can give the low-frequency dispersion of the collective spin waves.
Some analysis of the poles is shown in S5 of Supplemental Material \cite{SM}.
We fix the interaction strength $U=8$ to ensure the altermagnetic order, and numerically solve two transverse propagating modes with opposite chirality, $\omega_R(\mathbf{q})$ and $\omega_L(\mathbf{q})$, corresponding to the RH and LH precessional directions, respectively.

\begin{figure}[htbp]
	\includegraphics[width=\columnwidth]{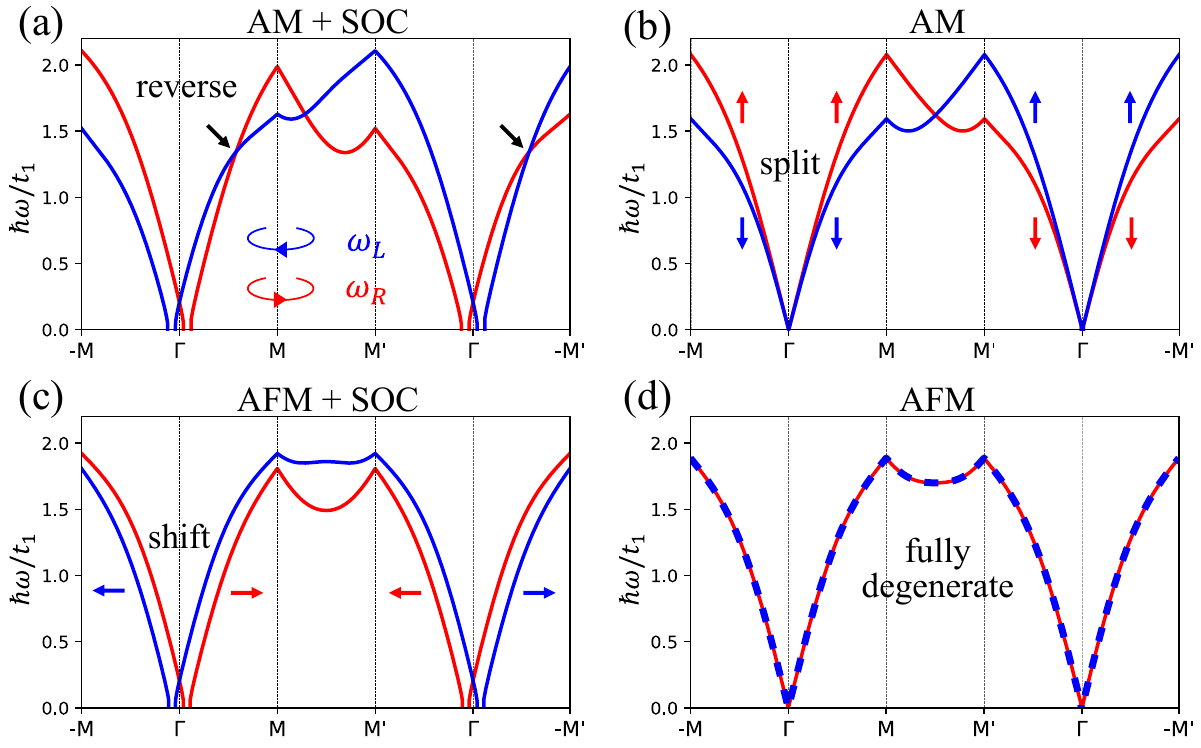}
	\caption{Spin-wave dispersion under different conditions. (a) $\delta=0.5,\lambda_R=0.1$. The two modes of the altermagnet (AM) with SOC exhibit anisotropic chirality splitting structure, which reverses its ordering in specific momentum directions. (b) $\delta=0.5,\lambda_R=0$. Without SOC, the chirality reversal disappears, and the structure recovers a simple splitting pattern with $C_{4z}\mathcal{T}$ symmetry. (c) $\delta=0,\lambda_R=0.1$. SOC shifts the LH and RH modes towards opposite directions, which may cooperate or compete with the altermagnetic splitting effect. (d) $\delta=0,\lambda_R=0$. In normal antiferromagnets (AFM), the two modes are fully degenerate. Other parameters are $t'=0.3,U=8$.}
	\label{pic_wq}
\end{figure}

Illustrated in Fig.~\ref{pic_wq}(a) is the dispersion of these two modes, which exhibits a highly anisotropic chirality-splitting structure.
In $\Gamma-(-M)$ and $\Gamma-M'$ ranges, the splitting is stable, while in $\Gamma-M$ and $\Gamma-(-M')$ ranges, a reversal of the splitting structure arises. These characteristics correspond to the electronic energy bands in Fig.~\ref{pic_band}(c).
Besides, several regions where we cannot find a real low-frequency solution appears around $\Gamma$, but they do not coincide for the two modes, thereby allowing nonreciprocal propagation.
The static solutions $\omega(\mathbf{q})=0$ lie at some $\mathbf{q}\neq0$ points, implying spatially modulated ground states.

In order to clarify the interplay between SOC and altermagnetism, we investigate the spin-wave dispersion under different conditions.
Fig.~\ref{pic_wq}(b) shows that when SOC is turned off, the splitting structure solely originating from the altermagnetic anisotropy exhibits the $C_{4z}\mathcal{T}$ symmetry, and is consistent with the z-polarized case in Refs.~\cite{smejkal_Chiral_2023, maier_Weakcoupling_2023, liu_Chiral_2024}, confirming a rotation symmetry of spin polarization orientation.
The static solution $\omega(\mathbf{q}=0)=0$ corresponds to the uniform ground state with the spontaneous $SU(2)$ symmetry breaking.
On the other hand, if we only keep SOC on and set $\delta=0$, the RH and LH modes experience opposite shift towards $+q_y$ and $-q_y$, respectively, as shown in Fig.~\ref{pic_wq}(c). They are also split as a consequence of the misplacement. The insoluble regions and spatially modulated ground states also originate from SOC.
Fig.~\ref{pic_wq}(d) shows that when the two origins both vanish, the chiral modes are fully degenerate.

Therefore, the chirality-splitting structure is a superposition of the splitting effect from altermagnetism and the shifting effect from SOC. The distinct symmetries make them cooperate or compete in different Brillouin zone regions.
Specifically, when $-\pi<k_x<0$, their splitting effects are opposite, with SOC dominating the long-wavelength range and altermagnetism dominating the short-wavelength range, which results in the chirality reversal.
The transition point highlights the parameters where the influence of SOC and altermagnetism cancels.
Note that such phenomenon only happens when spin polarization has in-plane components. For perpendicular polarization, the shifting effect of SOC is absent, thus only altermagnetic splitting effect remains, which is demonstrated in S6 of Supplemental Material \cite{SM}.

\begin{figure}[htbp]
	\includegraphics[width=\columnwidth]{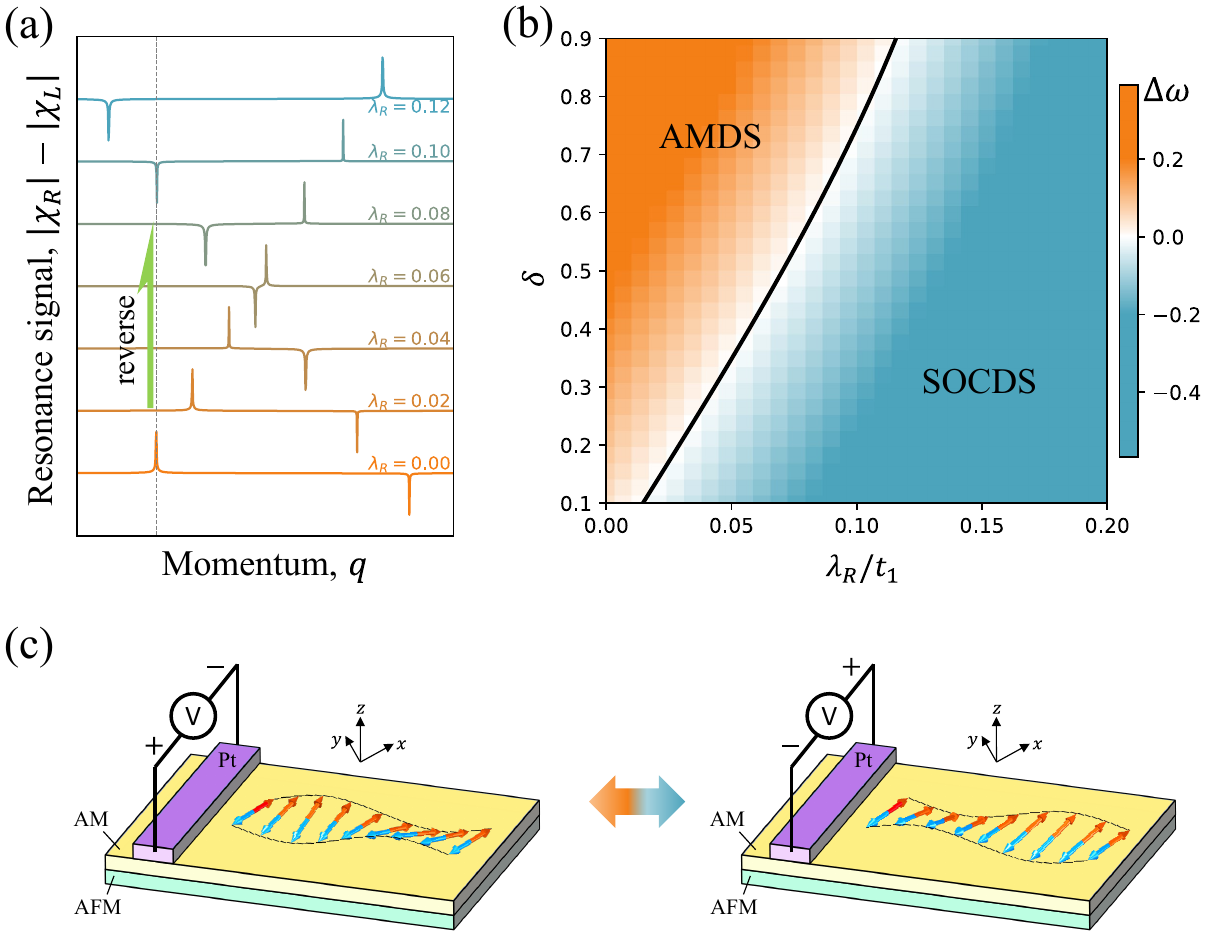}
	\caption{(a) Resonance signals (with offsets) for different SOC strength $\lambda_R$, and the signs represent chirality. With the increasing of $\lambda_R$, the peaks with opposite chirality move closer and finally swap positions, indicating the crossover from altermagnetism-dominated splitting (AMDS) to SOC-dominated splitting (SOCDS) structure. The vertical dashed line demonstrates that at fixed momentum, the signal reverses its sign. (b) Chirality splitting phase diagram of anisotropy and SOC strength. (c) Schematic illustration of the experimental design. The chirality of spin waves in the altermagnet/antiferromagnet (AM/AFM) bilayer is detected by the voltage in a platinum stripe and opposite chiralities generates different polarities.}
	\label{pic_switch}
\end{figure}

Since SOC and altermagnetism produce opposite chirality splitting structures in specific momentum directions, their competition offers an approach to switch the chirality of spin waves with fixed resonance conditions.
To detect the resonant chiral modes, we introduce a normalized quantity $|\operatorname{Re}\chi^{RPA}_R|-|\operatorname{Re}\chi^{RPA}_L|$ as the resonance signal.
As shown in Fig.~\ref{pic_switch}(a), the two peaks with opposite chirality gradually swap positions with the increasing of SOC strength, indicating the crossover from altermagnetism- to SOC-dominated splitting structure.
Crucially, if the excitation momentum $\mathbf{q}$ is fixed, the resonance signal will reverse its sign by tuning SOC.
We also give the chirality splitting phase diagram in Fig.~\ref{pic_switch}(b) with the characteristic parameter $\Delta\omega=\omega_R-\omega_L$, where we choose $\mathbf{q}=(-\pi/4,\pi/4)$ as a representative. The splitting structures in the two regions are dominated by SOC and altermagnetism, respectively. Tuning the parameters across the boundary may give us an opportunity to switch the chirality.

\textit{Experimental Implementation---}\noindent
To validate the proposed SOC-driven chirality switching, we design an experiment platform in Fig.~\ref{pic_switch}(c) comprising an altermagnetic thin film heterostructured with an antiferromagnetic substrate, where a lateral electrical current can induce tunable Rashba SOC \cite{zelezny_Relativistic_2014, manchon_Currentinduced_2019}.
The chirality of spin waves is detected by a platinum stripe on the top of the altermagnet, where the inverse spin Hall effect (ISHE) \cite{balinskiy_Spin_2021, li_Spin_2020, sheng_Control_2025, vaidya_Subterahertz_2020} converts opposite chiralities into opposite voltage polarities.
Initial resonant excitation by a terahertz irradiation with proper frequency generates a specific spin-wave mode \cite{li_Spin_2020, hortensius_Coherent_2021, vaidya_Subterahertz_2020}, producing a characteristic ISHE voltage signal.
When the SOC strength is enhanced to a proper value, a reversed voltage polarity will be detected, indicating the excitation of the other chirality, similar to the signal in Fig.~\ref{pic_switch}(a).
Alternatively, chirality-resolved dispersion relations can be directly probed using magneto-optic Kerr effect microscopy \cite{hayashi_Observation_2023, che_Discovery_2025}.
The demonstrated chirality-switching capability further can be used to construct chirality logic gates.

\textit{Conclusion---}\noindent
We have proposed a mechanism for electrically switching the spin-wave chirality in altermagnets through the modulation of SOC, instead of changing external fields, which is enabled by the competition between SOC and altermagnetism.
Our approach offers precise control and enhanced integrability over chirality operation, which paves the way for constructing chiral logic circuits \cite{khitun_Magnonic_2010, nikolaev_Operation_2023} and may accelerate the development of chirality-based computing devices \cite{chen_Reconfigurable_2021, bensmann_Dispersiontunable_2025}.
Moreover, our findings provide new insights into the interplay between nonrelativistic splitting effects and relativistic SOC, opening opportunities for discovering emergent phenomena arising form the synergy of these effects in quantum materials.


\begin{acknowledgments}
\textit{Acknowledgments---}\noindent
W. T. Li and W. M. Liu are supported by the National Key R\&D Program of China under Grants No. 2021YFA0718300, 2024YFF0726700 and 2021YFA1400900,  NSFC under Grants No. 12334012, 12234012, 52327808 and 62575314, and the Space Application System of China Manned Space Program.
J. T. Wang acknowledges the financial support from the NSFC under Grants No. 92263202 and 12374020, the National Key R\&D Program of China under Grant No. 2020YFA0711502, and the Strategic Priority Research Program of the Chinese Academy of Sciences under Grant No. XDB33000000.
\end{acknowledgments}

\bibliography{references.bib}

\end{document}